Enhanced superconducting properties of pre-doped B powder type MgB$_2$ strands

M.A. Susner, Y. Yang, M.D. Sumption, and E.W. Collings
*CSMM, MSE, The Ohio State University, Columbus, OH 43210, USA*

M.A. Rindfleisch, M.J. Tomsic
*Hyper Tech Research, Inc., 539 Industrial Mile Rd. Columbus, OH 43228, USA*

J.V. Marzik
*Specialty Materials, Inc., 1449 Middlesex Street, Lowell, MA 01851, USA*

Conventional doping methods that directly add C or a C-bearing species to Mg+B powder have the disadvantage of adding C inhomogeneously, yielding either under-reacted regions or blocking phases. Pre-doped B powder provides a more homogeneous distribution of the C dopant in MgB$_2$. Powders containing varying amounts of C were used to produce *in-situ* MgB$_2$ strands which showed high values of transport $J_c$ ($10^4$ A/cm$^2$ at 13.3T). Compared to SiC-added and malic acid-treated strands the pre-doped MgB$_2$ showed both higher values of $B_{irr}$ and transport $J_c$, indicating that the pre-doping of B leads to more efficient C substitution into the B-sublattice.

Interest in $MgB_2$ as a superconducting material has been substantial, both because of practical considerations such as its relative ease of fabrication and mid-range $T_c$ [1],[2], and more fundamental considerations which include its two-band nature [3]. Relevant to both of these categories, dopants which increase the irreversibility field, $B_{irr}$, and the upper critical field, $B_{c2}$, are subjects of particular interest, especially as the high values of $B_{irr}$ and $B_{c2}$ needed for applications are enabled by the two-band nature of the superconductor [3]. By far the most effective method of increasing the critical fields of $MgB_2$ is through addition of an element or compound that will lead to C-substitution into the B sub-lattice, producing disorder, thereby increasing $B_{c2}$, and hence the high field critical current density, $J_c$ [4],[5]. This type of substitution also has the benefit of decreasing the critical field anisotropy [6]. The drop in $T_c$ that is associated with C-doping-induced disorder is well known, but is a minor drawback when compared to the benefits of the resulting increases in $B_{c2}$ and $B_{irr}$.

Two main categories of C-doping exist for the *in situ* $MgB_2$ superconductor. The first may be called *indirect addition*. Here a C-containing species is added that decomposes before or during the $MgB_2$ reaction to form free C that then substitutes into the B sublattice. However, this doping route, by its very definition, inevitably adds impurities. By far the most common example of such a C-containing species is SiC which during decomposition yields a by-product, $Mg_2Si$, whose presence decreases the electrical connectivity between $MgB_2$ grains [7]. If we consider the relatively low reaction temperatures and times used in heat treating *in situ* wires, then we can assume that the diffusion of C into the B-sublattice is rather slow. As a result, the homogeneity of doping is determined by the distance between discrete dopant additions. We have already noted that C-doping by SiC results in a "smearing" of the superconducting heat capacity anomaly [8], which indicates that $T_c$ is distributed over a wide range of temperatures indicating that the level of doping in the superconducting strands is not uniform, resulting in chemical inhomogeneity.

The second category, *direct addition*, involves adding C to the Mg + B precursor powder mixture. Arguably the most successful example of this technique, at least in terms of resultant transport properties, is the malic-acid (MA) treatment, where MA is melted together with the precursor B at ~150°C and subsequently pyrolyzed such that only C remains surrounding the B agglomerates. The homogeneity of the dopant distribution should therefore be more uniform than with the *indirect addition* method. However, because the reacted $MgB_2$ grain size is, in the case of the *in-situ* method, strongly correlated with the operative length scales in the starting B powder (several micrometers down to less than 50 nm depending upon the type and provenance of the starting B powders), one now must consider the homogeneity of C-doping to be limited not by the size (spacing) of the dopant as is the case of SiC doping but by the size of the precursor B. Because the C is coating the precursor B, one would assume that, at least for the largest agglomerates, the C content on the outside of the final $MgB_2$ grain will be higher than at the center. In addition, with the MA-treatment, it is seen that the C-coating of the B can serve to prevent Mg vapor from fully reaching some B, resulting in the presence of B-rich regions that remain even after aggressive heat treatments of 700°C and 8 hours [9]. In the doping of *ex-situ* $MgB_2$ strands which will rely on C diffusion into the already formed (and typically larger grained) $MgB_2$, it is evident that the disadvantages of both *direct* and *indirect* C additions will be exaggerated.

Neither *direct* nor *indirect additions* will result in homogeneous doping unless the $MgB_2$ is heat treated at very high temperatures for long times. Consequently to ensure a more homogeneous distribution of C in $MgB_2$ at ordinary temperatues it is necessary to pre-dope the B powder.

For *in-situ* based $MgB_2$, smaller B powder sizes lead to higher $J_c$s, presumably due to finer grain sizes in the final $MgB_2$ [10]. The most consistent method of producing pure B is through plasma spray synthesis in which $BCl_3$ gas is reduced by $H_2$ during the injection of both gasses into an RF plasma [11-13], The resultant B is mostly amorphous with some nanocrystalline material. The plasma synthesis route also offers an opportunity to homogeneously dope with C through supplementing the gas stream with a hydrocarbon gas, e.g. $CH_4$, thus providing nanocrystalline grains or amorphous powders which contain elemental C within the B powders themselves. Such C-containing B powders, as manufactured by Specialty Materials Inc (SMI), were obtained with a variety of C-doping levels enabling monofilamentary C-doped $MgB_2$ *in-situ* PIT strands to be manufactured.

Monofilamentary powder-in-tube (PIT) strands 0.83 mm in diameter with a Nb chemical barrier and a monel outer sheath ($MgB_2$/Nb/monel) were manufactured by Hyper Tech Research, Inc. using a previously described process [14-16]. The starting powders consisted of Mg (99.9%, 20-25 μm) and the SMI-C-doped B. Using a LECO CS600 inorganic combustion analyzer, the final B powders were found to have C contents of 2.08 ± 0.05 wt% C, 3.47 ± 0.03 wt% C, 5.22 ± 0.26 wt% C, and 7.15 ± 0.11 wt% C. Table I lists these values together with the doping levels, in mol%, of the final $MgB_2$ compound. Straight samples of the drawn wire ~20 cm in length were reacted under flowing Ar gas for 20 mins. at 700°C.

Four-point transport $J_c$ measurements were made on short samples ~ 3 cm long in fields of up to 14 T applied transversely to the strand. A voltage tap separation of 5 mm was used; the $J_c$ criterion was 1 μV/cm. SEM micrographs were taken on polished longitudinal cross-sections of the strands using back-scatter mode on a Phillips Sirion SEM with a FEG electron source.

For each SMI-C-doped sample resistivity measurements as function of temperature, $R(T)$, through the superconducting transition in transverse magnetic fields of from 0.01 to 14 T were made on samples 8 mm long using a Quantum Design Model-6000 physical property measurement system (PPMS). Samples were chemically stripped of the Ni-Cu outer sheath to avoid any effect on the measurement from this ferromagnetic material. After re-arranging the $R(T,B)$ data, upper critical field and irreversibility field temperature dependencies, $B_{c2}(T)$ and $B_{irr}(T)$, were then extracted based on the 10% and 90% resistive-transition points, Figure 1; $\Delta T = B_{c2}(T)-B_{irr}(T)$, was obtained as a function of field. As discussed in [17] there are several contributors to $\Delta T$ such as: sample inhomogeneity or internal strain, thermally activated flux flow, and perhaps most importantly the Eisterer mechanism [18] which attributes the resistive transition to current percolation through a random mixture of grains whose anisotropic $B_{c2}$s are parameterized by $\gamma = B_{c2}^{\prime\prime}/B_{c2}^{\perp}$: According to Eisterer

$$\Delta T = \left(1-\sqrt{(\gamma^2-1)p_c^2+1}\right)B_0 \Big/ \left(\partial B_{c2}/\partial T\right) \qquad (1)$$

in which $\partial B_{c2}/\partial T$ is the critical field temperature dependence (Figure 1) and $B_o$ denotes the external field. Based on this model and assuming 0.3 for the percolation threshold, $p_c$, as previously used in calculations of transport current percolation through porous $MgB_2$ [19], we are able to show how critical field anisotropy depends of the level of C doping in our SMI-C samples, Figure 1(inset).

Figure 2 shows the 4.2 K transport $J_c$s as a function of magnetic field for various doping levels. Critical current densities of $10^4$ A/cm$^2$ were achieved at 6.4 T for the undoped $MgB_2$-00, 9.5 T for $MgB_2$-01, 10.9 T for $MgB_2$-02, 13.3 T for $MgB_2$-03, and 12.8 T for $MgB_2$-04. The highest transport values were seen for MgB$_2$-03, e.g. $10^4$ A/cm$^2$ at 13.3 T. MgB$_2$-04 exhibits lower $J_c$s, e.g. $10^4$ A/cm$^2$ at 12.8 T, indicating that the C has exceeded its solubility limit [20] and is presumably collecting at the grain boundaries or in large clusters, thereby reducing connectivity.

Figure 3 presents the relationship between the analyzed C content of SMI-C-doped MgB$_2$ and the 4.2 K $B_{irr}$ calculated using the 100 A/cm$^2$ criterion. A comparison to both SiC- added and MA treated samples is also presented. For these samples the C contents are based upon assuming complete disassociation of the SiC or LECO analysis of the MA-treated B powder. The pre-doped MgB$_2$ samples have the highest values of $B_{irr}$ at every doping level. The MA-treated MgB$_2$ is relatively close to the pre-doped (SMI-C) line, suggesting that for this directly C-doped sample, substitution of C into the B sublattice is nearly complete. The microstructure typical of the MA-treatment, with areas of higher boride phases is also shown (Figure 3d).

Indirect C-addition via the 200 nm SiC-added MgB$_2$ sample leads to the greatest departure of $B_{irr}$ from the SMI-C-doped line. This is a result of the relatively large size of the SiC particles and hence, for a given concentration, the relatively large distance between them which leads to a long C diffusion length. By contrast, the 15 nm SiC-added MgB$_2$ sample shows an increased value of $B_{irr}$ for the same amount of SiC present, suggesting that the shorter distance between SiC particles resulting from the smaller particle size decreases the diffusion length, allowing more C to substitute into the B-sublattice thereby increasing $B_{irr}$. Also contributing to the departures of the SiC-doped $B_{irr}$s from the SMI-C-doped line is incomplete decomposition of the SiC particles as evidenced by the presence of unreacted SiC (Figure 3c). Again, it is seen that MgB$_2$-04 does not follow the trend of the other pre-doped samples, with a visible plateau in $B_{irr}$ at the expected doping level of 4.32 mol% C, once again suggesting that C has exceeded its solubility limit [20] in the B sublattice (Figure 3b).

In summary, Pre-doped B powders with various levels of C-doping were used to manufacture $MgB_2$ superconducting strands. These strands were found to have high 4.2 K transport $J_c$ values, with sample $MgB_2$-03 carrying $10^4$ A/cm$^2$ at 13.3 T (154 kA/cm$^2$ at 6 T) and a $B_{irr}$ of 23.5 T (as compared to 20.3 T and 16.1 T for 15 and 200 nm SiC additions with equivalent C contributions). An increase of the C content to 4.32%, marginally increasing $B_{irr}$ to 24.4 T, but had a deleterious effect on $J_c$, then characterized by $10^4$ A/cm$^2$ at 12.8 T. This phenomenon was attributed to dopant saturation where C formed clusters outside of the $MgB_2$ grains, serving to reduce electrical connectivity. The pre-doping mechanism was found to be a more efficient doping route than either the direct MA-treatment or indirect SiC addition, with higher critical fields reached for a given level of addition.

This work was supported by the United States Department of Energy, HEP grant DE-FG02-95ER40900.

Table I. List of Sample Parameters

| Sample Name | HTR Tracking # | SC fraction | wt% C in starting B powder | mol% C in final $MgB_2$ |
|---|---|---|---|---|
| $MgB_2$-00 | 2134 | 0.252 | 0 | 0 |
| $MgB_2$-01 | 1990 | 0.115 | 2.08 ± 0.05 | 1.25 ± 0.03 |
| $MgB_2$-02 | 1650 | 0.152 | 3.47 ± 0.03 | 2.09 ± 0.02 |
| $MgB_2$-03 | 1991 | 0.131 | 5.22 ± 0.26 | 3.15 ± 0.16 |
| $MgB_2$-04 | 1952 | 0.183 | 7.15 ± 0.11 | 4.32 ± 0.07 |

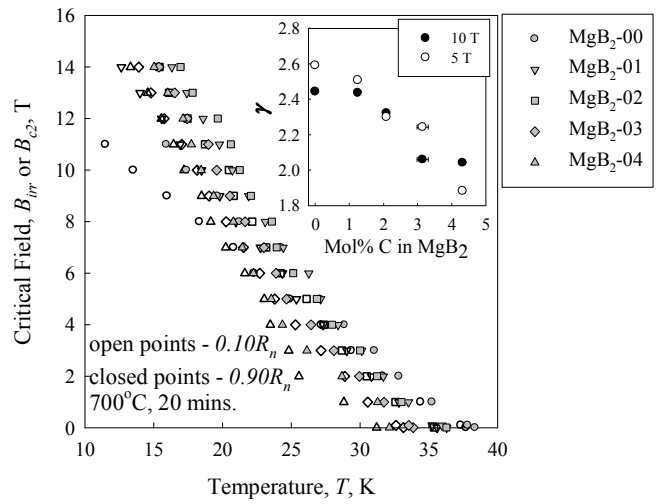

Figure 1. Critical fields, $B_{irr}$ and $B_{c2}$, vs. temperature for various levels of C-doping. (Inset) Anisotropy ratio $\gamma$ vs. mol%C at 5 T and 10 T.

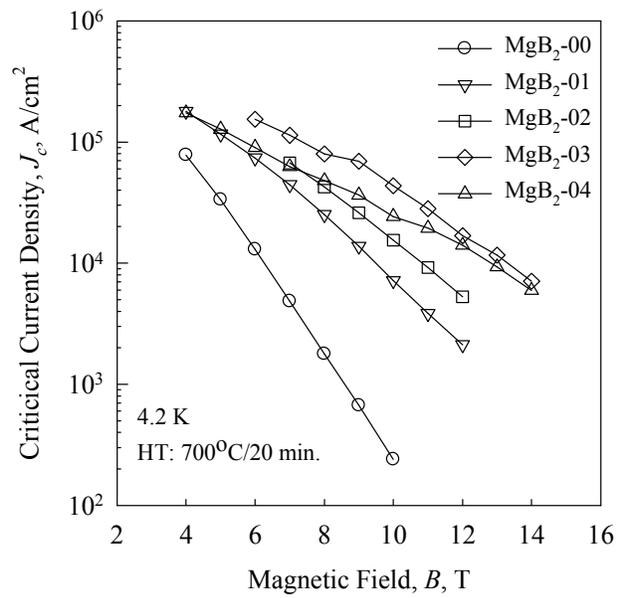

Figures 2. 4.2 K Transport $J_c$ for increasingly C-doped $MgB_2$ strands.

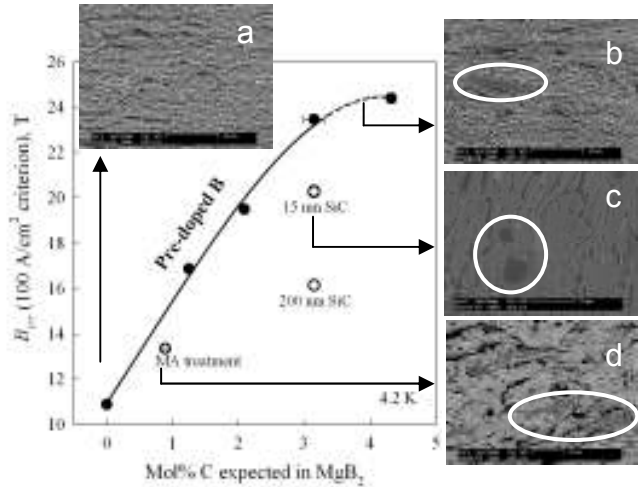

Figure 3. $B_{irr}$ (100 A/cm$^2$ criterion) as function of expected C-doping for pre-doped MgB$_2$, SiC-added MgB$_2$, and MA-treated MgB$_2$. The microstructures for each method of C-doping are presented for comparison. (a) the undoped MgB$_2$ shows a porous fibrous microstructure with no discernable inhomogeneities; (b) the heavily pre-doped MgB$_2$ shows some areas of B- and C-rich impurity phases indicating C saturation; (c) the SiC-added MgB$_2$ shows SiC agglomerations after HT; (d) the MA-treated MgB$_2$ shows B-rich regions.